# WIRELESS SENSOR NETWORK TECHNOLOGY FOR MONITORING MOISTURE CONTENT OF WOOD

**Ivan Araquistain, Jose Miguel Abascal, Oriol Munne**
Tecnalia Research and Innovation
Polígono Lasao, Area Anardi
nº 5 20730 (Azpeitia)
GUIPUZCOA (SPAIN)
TELEPHONE: +34 946 430 850

Iván Araquistain, Jose Miguel Abascal, Oriol Munne

## Abstract

Leaks represent a very important hazard for the buildings and they can affect all sorts of building materials and specially wood due to its hygroscopic properties. Excessive moisture content can affect in a negative way building processes such as the installation of wooden floors or the use of wood as a structural material.

Moisture meters can provide prompt and non-destructive determination of wood moisture, and as such are among the most useful tools available to wood products manufacturers and scientists. However, a continuous monitoring system is needed in order to avoid excessive moisture content which can damage wooden floors as well as structural wood. Data and procedures are presented in order to develop a suitable monitoring tool based on wireless sensor networks to provide an electronic tool of active security both for the installation of wooden floors and for the proper maintenance of existent buildings which have a timber structure.

# 1   Introduction

Nowadays, one would think that, after years of massive concrete and steel based construction in European countries, there are not many wood structure buildings left to be refurbished except for some palaces or cathedrals. However, if we have a look at the old part of many cities, we will realize that still most of the buildings have a wood structure. [1]

Because of their historical value, or preferred location, urban plans prohibit their demolition, they grant a permission for refurbishment works which clearly indicate that moisture is one of the key hazards for the decay of wooden structures.

**Moisture and sanitary status of the wood structure**

The determination of sanitary conditions is to detect, identify and assess damage from biotic and abiotic origin that has affected the structure.

According to a survey by Bureau Securitas carried out for over 10,000 building damages, 23% of them had their origin in the rain, snow or wind, 8% were related to corrosion, 49% were due to thermal and hygrometric variations and 20% were due to other causes. That is to say, a great amount of damages are linked directly or indirectly to leaks and moisture.

Main abiotic agents that can degrade a wooden structure are [2]:

- Atmospheric agents: solar radiation and humidity changes

- Chemicals: bleach, detergents, acids...

- Mechanical: abrasion, acts of man...

- Fire

Main biotic agents that can cause decay are:

- Mushrooms, xylophages, fungi and decay fungi chromogens

- Marine borers: mollusks and crustaceans

- Insect borers: larval cycle insects (Lepidoptera, Hymenoptera and Coleoptera) and social insects (termites)

Moisture content is a key factor in the presence of organisms attacking wood, especially fungi which usually starts to attack wood above the 18-20% humidity.

**Moisture related problems in wooden floors**

Moreover, only in Spain, in 2004 the sector of wooden parquet floors moved 263,284 million Euros. More than 8 million square meters of this material were consumed, 66% of which was produced in Spain, while the rest was imported from other countries.

Only in Spain, it can be calculated that every year, 15 million Euros are lost due to problems with moisture in wooden floors. This figures would be much higher is we would take into account other countries.

**Hypotheses and objectives of the project**

Willing to approach this problem, the project assumes that it would be possible to develop a wood moisture monitoring system both for wooden floors and structures based in the concept of early detection and prediction of defects in building processes so as to avoid damages in new buildings as well as in old ones.

It is well known that moisture affects the living conditions and health of the inhabitants of a building. This document will explain how in the future Wireless Sensor Network technology will be able to enhance the maintenance of a wood structure building while reducing costs and environmental impact.

## 2  Current moisture measuring techniques applied to wood

The building inspection is aimed at detecting construction defects in the building that may affect the wood structure, determining the health status of the structure including characterization, evaluation and classification of each of the parts constituting the structure. Excessive moisture is one of the main biotic and abiotic damage causes. It is also essential to know the moisture content of wood material before the installation of wooden floors to avoid dimensional stability problems.

## 2.1 Moisture measuring equipment

*-Wood hydrometer:* This equipment is intended to determine on-site, quickly and easily, the wood moisture content. Most frequently, it is based on the electrical conductivity of wood depending on moisture content. See equation (1), where R is the electrical resistance in Ohms and M the moisture content in %.

$$Log\ [log(R)\ -4] = 1.009-0.0322M \qquad (1)$$

However, Non-contact wood hygrometers based on capacitive moisture sensing are now gaining market as they are easier to use.

Moisture content is a key factor in the presence of wood xylophagous, especially for fungus, which generally starts to attack wood when moisture content is over 18% - 20%.

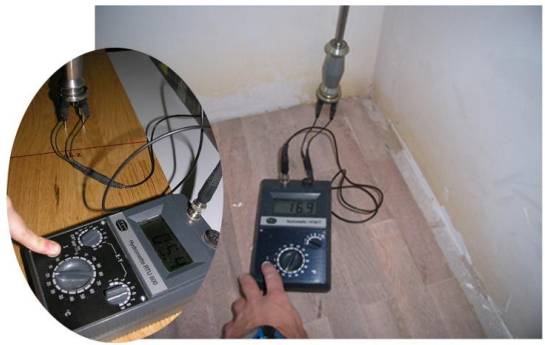

Figure 1: Wood moisture measurement

*-Infrared Thermography:* Thermography is the use of an infrared imaging and measurement camera to "see" and "measure" thermal energy emitted from an object as it detects IR heat and converts it to electrical signals that are displayed on video screen. These cameras have shown great potential as Non-Destructive evaluation tools. Infrared thermography is becoming a technique that is being used to detect moisture and decay in wood structure buildings. Many other techniques are cited by authors [3], [4],[5].

## 3 Material and Methods

This project aimed to create a real time wireless monitoring system of the moisture content in any wood based product to provide an electronic tool of active security both for the installation of wooden floors and for the proper maintenance of existent buildings which have a timber structure. Therefore it was essential to establish clear thresholds to activate alarms that indicate that a problem may occur. In order to establish threshold levels for the products of the companies participating the project the dimensional stability of parquet floor samples was tested in a climatic chamber and also a threshold moisture level was established for structural wood.

### 3.1 Equipment requirement specification

#### 3.1.1 Wooden floors

For testing dimensional stability of the oak (*Quercus Robur*) parquet, sized 2190x210x14mm the following procedure was followed. First, upon receipt of the material to be tested, initial moisture was measured both by a calibrated moisture measuring equipment (GANN RTU 600 hygrometer) and by a drying oven according to the procedure described in UNE EN13183-1 (2004) and initial dimensions were measured. Initial moisture content was of 6% with little variation.

Then, 12 specimens were introduced in the climatic chamber under conditions of 85% humidity and 23 ° C temperature. 10 of these specimens were used for measurements of curl and curvature of the face, while the other two were used to obtain samples that permitted the exact measurement of moisture content by oven-drying method. 5 of the specimens were placed directly on the shelves of the climatic chamber, while the other 5 were placed on battens.

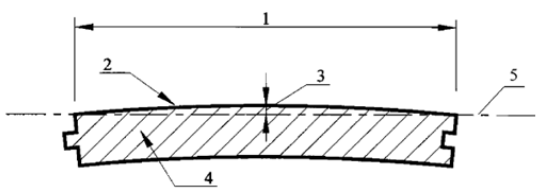
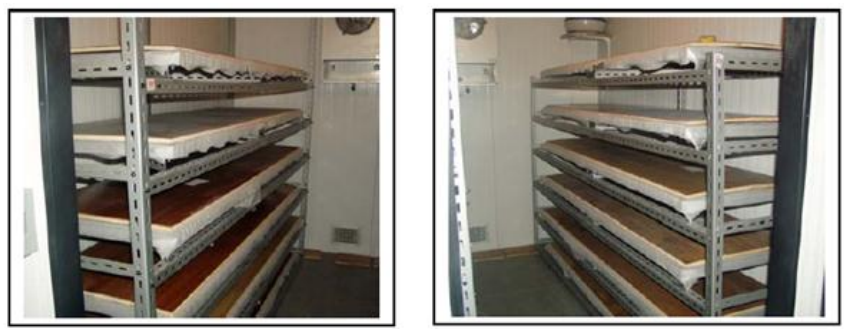

Figure 2: Wood parquet samples in the climatic chamber

Successive measurements were made to determine the curvature of the face and the curling of the specimens as long as they gained moisture content. All of the 10 samples exceeded the specifications of face curling after the exposure at the climatic chamber.

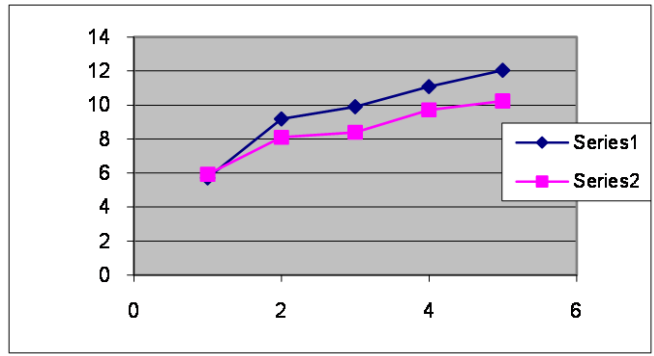

x-axis: Successive measuremets
y-axis: Misture content in %
Series 1 corresponds to: Specimens on battens
Series 2 corresponds to: Specimens directly on shelves

Graph 1: Moisture content of parquet samples

It is noted in Graph 1 that the moisture increasing of parquet placed on battens is faster than that of the ones placed directly on the shelves. Logical outcome, as in the case of the specimens on battens, moist air is in contact with both sides of the specimen.

The following results were obtained:

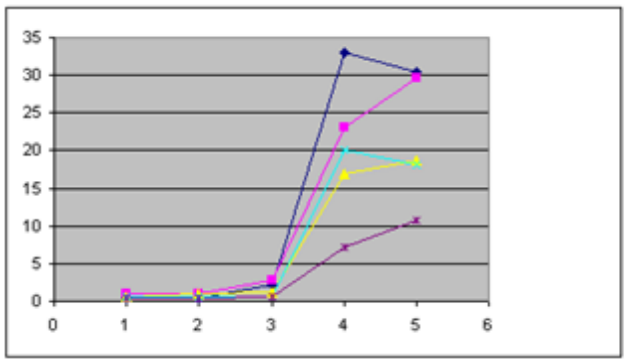 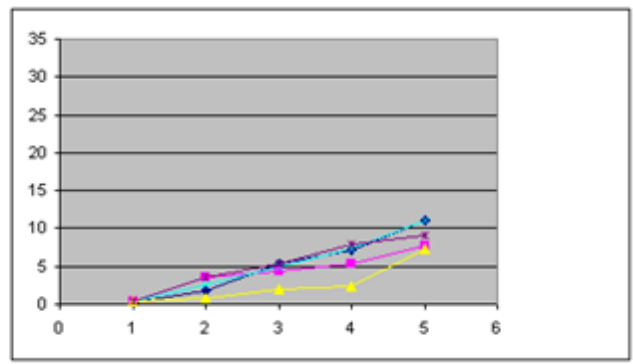

Graph 2: Face curvature of parquet specimens on batten (left) and on shelves (right)
x-axis: Successive measurements
y-axis: Face curvature in mm

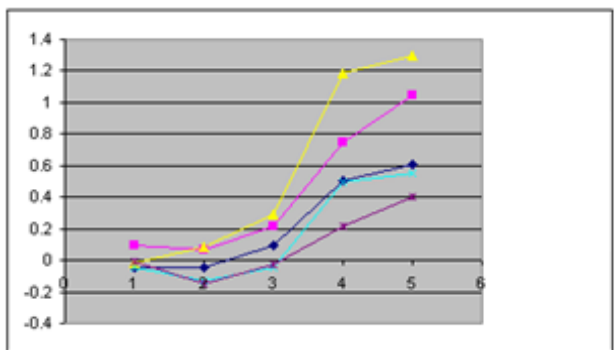 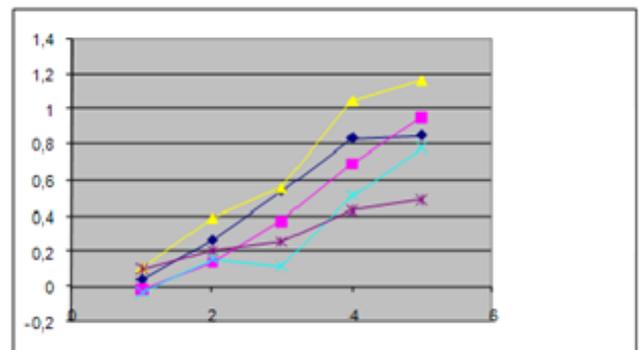

Graph 3: Curl of parquet specimens on batten (left) and on shelves (right)
x-axis: Successive measurements
y-axis: Curl in mm

\* Note: The maximum allowed curl according to the standard (UNE EN 13489:2002) is 0.2% of the width. In this case corresponds to 0.42 mm. There is no specification for the curvature of face.

As Graph 3 shows, all specimens exceeded curl specifications which could constitute a basis to reject the material. This defect was evident when moisture content reached from initial 6% to 10-11% of moisture content.

### 3.1.2 Structural wood

Excessive moisture content (over 20%) can affect in a negative way structural wood, that is why, long term monitoring can enhance wood protection limiting moisture threshold to trigger an alarm to 20% of moisture content. Optimum moisture content for the growth of xylophages is between 25% and 55%.

### 3.2 Moisture measuring technology used in the wireless sensor

As a result of the tests for wooden floors, it was decided that the equipment should send an alarm signal whenever moisture increased over 3 points (3%) from initial moisture content. This alarm should serve as a warning to take immediate corrective action, considering that a further moisture increase can lead to permanent damages. For wooden structures a 20% moisture limit was decided.

Once the limits where established several electronic prototypes were developed in order to produce a wireless sensor capable of measuring moisture content out of 10 independent points, having a web application were the online monitoring results could be followed.

Moisture measuring technology used in the wireless sensor was based on the electrical resistance of wood, as it varies with the moisture content of it and to a lower extent with the temperature. In addition to this it depends on the wood species.

In dry condition, wood is a very good isolator for electricity, with an approximate resistance of (1.000 MΩ). However, with a 30% of moisture content, the electrical resistance of wood only reaches about (200mΩ). Therefore, it can be stated that the sensibility of the moisture measuring equipments is much higher when it comes to measuring low moisture rates than when it comes to measuring high ones. In spite of this fact, for detecting possible problems in wood structures, it is enough with measuring up to the 25% of moisture content with a correct accuracy (threshold for alarms is at 20%), even if for greater moisture contents, the accuracy level decreases.

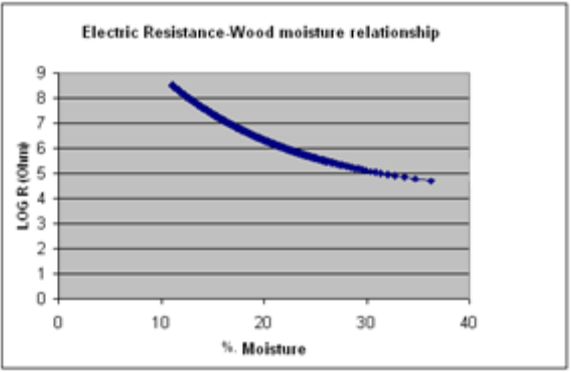

x-axis: Moisture content in %
y-axis: Log R (where R is resistivity in Ohms)

Graph 4: Electric resistance vs. wood moisture

## 3. 3 Brief overview of the electronic circuit and web application developed

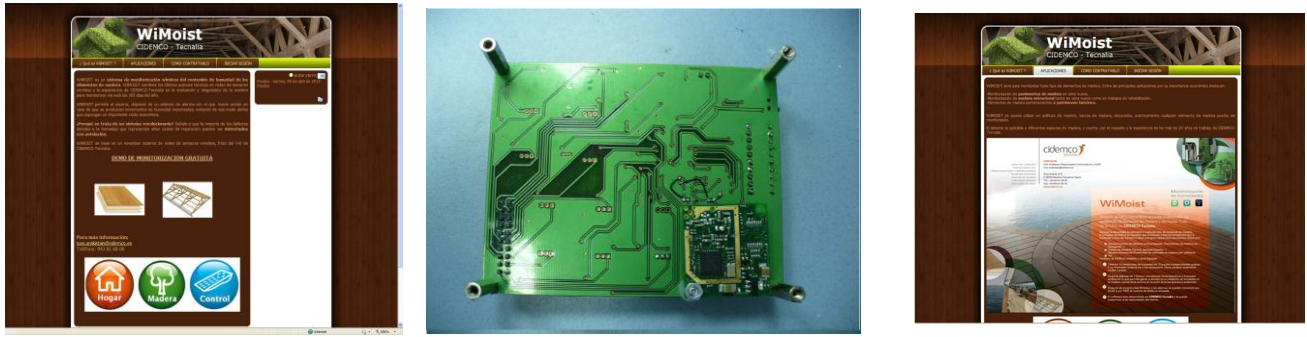

Figure 3: Electronic measuring circuit prototype and web application

Ultra low power consum______________________ is currently being developed. This technology is specially helpful for application where a low data transmission rate is needed. For example, for measuring moisture content of wood, a couple of measurements per day would be enough.

The great advantage of this technology is that many cost effective and little sized sensors can be used to monitor an environment in a PC which is located anywhere in the world. This sensors have ultra low power consumption taking advantage of special data transfer protocols such as (Zigbee). As a consequence, the goal for the future is to develop cost effective sensors with a battery life of about 5 years with a daily use of two measurements per day.

The web application enables the user to upload photographs from the areas where the sensors have been located. The application is organized by the number of floors of the building, which can be created and managed by the user.

The developed software has an implemented function to calculate moisture tendency, depending on current equilibrium moisture of wood and environmental moisture and temperature which will lead wood moisture to increase or decrease. This function is particularly useful for wooden floors, which are more likely to suffer from severe damages related to dimensional stability problems.

## 4 Results and Discussion

In conclusion we found that, having controlled the production process so that the moisture out of it is 6% (which was observed in the parquet tested although the standard (UNE 56810:2004) allows to install parquet humidity is between 5% and 9%), the equipment should give an alarm signal whenever more than 3 points of increased humidity (see Graphs 2 and 3). This alarm should serve as warning to take immediate corrective actions, taking into account that there was 1% of admissible increase in moisture wood before going outside the standard.

Regarding the development of the wireless sensor system and web based software most significant innovations regarding the state of art were the development of a moisture tendency algorithm to predict problems of dimensional stability in wooden floors, and the development of a wireless sensor with the ability to measure moisture from 10 independent points (see Figure 3). The prototype equipment was tested at laboratory level although some accuracy problems were encountered for measuring moisture from distant locations.

## 5 Conclusions

About 40% of Spanish buildings are over 50 years old, many of them with a wood structure, apart from infrastructure. Restoring them represents a big sum of money, difficult to reach due to the economical crisis.

That is why a special effort is needed to improve current inspection and maintenance procedures, so as to be cost-efficient at restoration works, maintaining as much as possible of our cultural heritage.

However, assessment is a challenging activity in need of inputs from Non Destructive Techniques and wireless monitoring to be used at full capacity. The future of construction and refurbishment is probably strongly linked to ICT development, which will provide new inspection techniques and procedures to monitor and ensure the quality of constructions.

## 6 Acknowledgements

In the project, three Spanish companies have taken part, two of them related to wood industry and an electronic circuit producer company while the project was co-funded by the Industry Ministry of Spain and the Basque Government which has also contributed for the European patent protection of the results.